# CONCEPTUALISING HEALTHCARE-SEEKING AS AN ACTIVITY TO EXPLAIN TECHNOLOGY USE - A CASE OF MHEALTH

Karen Sowon, University of Cape Town, kc.sowon@uct.ac.za

Wallace Chigona, University of Cape Town, Wallace.chigona@gmail.com

**Abstract**: The purpose of this paper is to engage with the Information Systems' contexts of use as a means to explain nuanced human-technology interaction. In this paper, we specifically propose the conceptualisation of healthcare-seeking as an activity to offer a richer explanation of technology utilisation. This is an interpretivist study drawing on Activity Theory to conceptualise healthcare-seeking as the minimum context needed to explicate use. A framework of the core aspects of AT is used to analyse one empirical mHealth case from a Kenyan context thus illustrating how AT can be applied to study technology use. The paper explicates technology use by explaining various utilisation behaviour that may emerge in a complex human-technology interaction context; ranging from a complex adoption process to mechanisms to determine continuance that differentiate trust in the intervention from trust in the information, and potential technology coping strategies. The paper is a novel attempt to operationalise AT to study technology use. It thus offers a broader explication of use while providing insights for design and implementation made possible by the conceptualisation of healthcare-seeking as an activity. Such insights may be useful in the design of patient-centred systems.

**Keywords**: Activity theory, Technology use, mHealth, Healthcare-seeking

## 1. INTRODUCTION

Context plays an important role in how a repertoire of Information Systems (IS) are used (Cecez-Kecmanovic et al., 2014), and is therefore critical in understanding individuals' technology use. Characteristics of the context may provide mechanisms for success or failure in users' interaction with innovations. Thus, taking a completely social or completely technical perspective to IS behaviour are extreme positions. Valuing the social and the material over the technology is problematic as it leads to some form of analytical fallacy; since technologies have been known to have agency (see Orlikowski, 2005). The nature of technological agency and whether it is delegated or not is a subject we do not address here.

Given the balancing act that is required to study the interaction between technology and the social without embracing technological determinism, taking the technology for granted, or allowing the technology to vanish from view, a robust approach is necessary. Such has been the argument behind using a socio-technical approach in IS studies. An example of such an approach is the perspective of studying technology in use (in-situ). Such an attempt would argue for example that a hammer in a mason's hand is epistemologically different from a hammer in a blacksmith's hand; all because the 'context of use' differs between these two environments. The argument, therefore, is that such an approach to analysing both the context and the technology offers a more nuanced understanding of the complex relationship between humans and technology such as technology use.

In IS, while there have been many grand theories that have been useful in understanding the use and adoption of technology, these theories have been criticised for being too techno-centric assigning all the change agency to the technology (Awa et al., 2016; Venkatesh et al., 2007). To address these concerns, IS researchers have had to turn to social theories to offer a better understanding. These theories provide the vocabulary to address the "*situated entanglement of the technology and the social*" (Orlikowski, 2005, p. 185). Thus, social theories allow the researcher to study the characteristics of the context including the social aspects that may influence how people interact with technology. Some of the commonly used theories include Actor Network Theory, Structuration Theory and Activity Theory (AT), which have been useful for explicating the broader relationship between technology and human activity (Karanasios & Allen, 2014; Klein & Myers, 2001). While Structuration Theory is beneficial for studying organisational IT use, ANT assigns equal agency between human and non-human actors (Leonardi, 2011). In other words, in ANT, there is symmetry of agency between the human actors (users) and non-human actors (technology artifacts). AT on the other hand acknowledges that while both the technology and the actor (user) have agency, the relationship is asymmetrical. Thus, AT provides a more compelling theory to study technology in use since we start off with the assumption that human agency has intentionality (Cecez-Kecmanovic et al., 2014). In other words, though human actors and technology have agency, "ultimately, people decide how they will respond to the technology" (Leonardi, 2011, p. 151). Hence AT's perspective is one of 'agentic people actively pursuing an objective, not passively accepting and using technology' (Blayone, 2019, p. 457).

AT has been embraced in fields related to IS to understand work processes and transformation. These fields include HCI (Bødker, 1996; Hautasaari, 2013), Information Systems Design (e.g. Mursu et al., 2006) and computer-supported collaborative work (e.g. Zurita & Nussbaum, 2007). This is reasonable since primarily, the strength of AT lies in its ability to theorise human doing as an activity and understand this activity through its different phases of existence; what is called the activity network. AT has also been extensively used in educational technology to understand how teachers and students use technology artifacts to achieve their teaching and learning goals (see Karakus, 2014). While a quick look at the literature seems to suggest that AT has gained traction, a deeper analysis indicates that the IS field may benefit from more studies that show how AT may be operationalised to study other IS phenomena. While having the concepts to frame IS research (Karanasios & Allen, 2014), the theory's application to IS research has remained abstract and relatively scarce in IS-related phenomena like use.

Technologies are continuously changing, and complexity in interaction is ever increasing with their varied application in almost every sphere of life. Of interest is the growth of innovations in healthcare especially as they are used by consumers to facilitate their healthcare-seeking. These include mHealth, which is of specific interest to this study. mHealth which has been defined as the use of mobile devices to deliver healthcare services has experienced growth in developing country contexts (Lee et al., 2016). The growth is largely attributed to the potential that this technology presents to increase healthcare delivery, bridge the health inequality gap, and hence help developing countries to faster address their health-related goals.

The healthcare-seeking context where such technologies are introduced is a complex environment. It represents a combination of multiple stakeholders that may have similar or conflicting objectives, and whose achievement may be mediated by a panoply of factors. Some of these critical factors include socio-cultural beliefs which in healthcare-seeking define the meaning of sickness and when one can or should not seek healthcare. As such, the use of technologies like mHealth in such a space is a phenomenon worthy of careful study to understand why, when, and how such interventions work when they do. Such studies will pave way for the scalability and replicability of interventions in other contexts since such innovations are not a 'one size fits all'. Since AT provides a way to theorize human activities and understand them within their context of use, we argue that studying use from the perspective of AT will prove useful. This study, therefore, seeks to answer the following question:

> *How can conceptualising healthcare-seeking as an activity be used to explain mHealth use?*

By studying technology in situ, proponents of such approaches argue that the examination of how humans interact with technology could inform human-centred IS design so that technology solutions are effective and appropriate (Bannon & Bødker, 1989; Benson et al., 2008; Karanasios, 2018). Hence, in answering this question using AT, we desire to show that we can generate insights for the design and implementation of technologies like mHealth. While we use the case of mHealth, we believe that the insights will be useful in varied contexts seeking to understand the use of consumer health information technology as well as other technology use contexts.

Using AT as the conceptual basis, this study analyses one healthcare-seeking empirical context where an understanding of technology use was sought. We adopt maternal health as an instrumental case for its nuanced rich context, which would better demonstrate all the concepts in an AT context.

The rest of the paper is organised as follows: The next section defines healthcare-seeking and introduces the AT concepts that are applied to conceptualise healthcare-seeking as an activity. The mHealth case study is then presented, and we thereafter demonstrate how the proposed conceptual framework can be applied, by using it to analyse the empirical case. We conclude by discussing the contribution of the paper and highlighting opportunities for further research.

## 2. CONCEPTUALISING HEALTHCARE-SEEKING AS AN ACTIVITY

### 2.1 What is Healthcare Seeking?

In health, patient health outcomes are not the sole result of efficacious health interventions (Alexander & Hearld, 2012). The impact of these interventions is mediated by the context. Commonly, socio-cultural factors may enable or impede the achievement of the desired health outcomes. These hindrances or enablers often manifest in healthcare-seeking behaviour which has been defined as the actions or inactions that individuals undertake for the purpose of finding a remedy in their perceived illness (Olenja, 2003). These decisions to seek and use healthcare may subsequently be surrounded by various beliefs of the nature, cause and meaning of sickness. These subsequently determine when, where, and how patients seek healthcare especially in medically pluralistic environments such as is common in Africa (Aikins, 2014). Thus, theorizing healthcare-seeking as an activity requires the chosen approach to have a means of capturing the nuances of the healthcare-seeking context, while being able to account for the goals of the healthcare-seeker when engaging in healthcare-seeking.

The next section presents the underpinning of AT, from which we derive a conceptual lens to examine healthcare-seeking as the total activity that provides the context to understand mHealth use. This lens is applied to analyse one empirical study in the later sections of the paper.

### 2.2 Introduction to Activity Theory

AT has its foundations in Russian psychology. "An activity is a form of doing [that is] directed to an object" (Kuutti, 1996, p. 27). The theory argues that human activities like learning or seeking healthcare are historically, culturally, and socially situated (Leontiev, 1978). *Activities* represent human agency in undertaking purposeful actions and consequently give meaning to human actions. This implies that human actions can only be understood within the context of the activity, which is the notion behind AT's principle of *'unity of consciousness'*. While AT has a rich genealogical history, time and space do not allow us to focus on those details in this paper. However, we choose to apply the second generation AT that was expanded and popularised by Engeström (Engeström, 2001), which recognised the biggest limitation of the first generation as being the focus on the individual. The addition of building blocks in the second generation like rules, community, and division of labour provides AT a further means to account for and understand context (Allen et al., 2011).

AT posits that with the help of a tool, a subject undertakes an activity, motivated by a particular motive/object (Allen et al., 2013). The *subject* is the individual 'whose agency is chosen as the point of view in the analysis' (Hsu et al., 2010, p. 1246) and who has the aim to transform the object into a desired outcome. A maternal health client, therefore, engaging in healthcare-seeking in the pursuit of a successful pregnancy, and whose utilisation of a health information technology artifact is being studied would be an apt representation of a 'subject'. The *motive/object* of a healthy pregnancy then would give meaning and direction to the actions or chains of actions (like Health Information Technology (HIT) use or lack of it thereof) which are carried out by the subject. This argument on the importance of the motive as being the reason for the activity is defined by AT's principle of *'object orientedness'*. The *Artifacts/Tools,* (e.g. mHealth) acting as resources for the subject in the pursuit of the motive (Yamagata-Lynch, 2010) would therefore enable or constrain the subject's potential to manipulate and transform the object of the activity. This principle of *'mediation'* is therefore critical in fully explicating artifact use within its activity context.

*Rules/Values.* Rules mediate the subject-community relationship. They are both implicit and explicit (Kuutti, 1996) and place limits on the activity being undertaken. A researcher applying AT may distinguish between the explicit norms (rules) and the implicit norms (values). For example, in a maternal healthcare-seeking context, the latter may include the healthcare system rules like the need for an appointment card that shows a record of all ANC appointments as a prerequisite for a mother to be admitted into a health facility for child delivery. Values on the other hand may include the culturally shared beliefs, norms, and values about pregnancy. For example, in some African societies, it is believed that the early periods of pregnancy are particularly susceptible to witchcraft (Nyemba-Mudenda & Chigona, 2018; Simkhada et al., 2008). In light of such beliefs, a woman may be compelled to keep the pregnancy secret to safeguard herself from people that could 'harm' the pregnancy. Openly disclosing one's pregnancy may also be associated with boastfulness which may result in embarrassment should one not carry the pregnancy to term (Pell et al., 2013). Pregnancy is also evidence of sexual activity, and consequently, women may be shy to disclose their pregnancy (Finlayson & Downe, 2013). All these cultural perceptions influence when and how women in developing countries engage in healthcare-seeking for example by starting ANC late in the third trimester only when the pregnancy is evident/showing (Simkhada et al., 2008; Uldbjerg et al., 2020; Wang et al., 2011).

*The Community.* The community is generally invested in the object of the activity system. Yamagata-Lynch (2010, p. 2) clarifies that the community "is the social group that the subject belongs to while engaged in an activity". The community may, therefore, refer to a community of practice, or a community of purpose. The former shares common ways of seeing and doing things (Hsu et al., 2010), while the latter connotes a shared motive, rather than practice. Since the maternal healthcare-seeking context reflects the coming together of stakeholders that share a motive, say the desire for a successful pregnancy, we hereafter adopt the term community of purpose (CoP) for its appropriateness to this research.

In a maternal healthcare-seeking context, the CoP may include the healthcare providers, as well as various other individuals that may have a stake in a pregnancy. For example, female kin remain crucial because pregnancy-related authority may be socially defined as belonging in the female domain (see Ensor & Cooper, 2004; Mumtaz & Salway, 2007). On the other hand, partners have been reported to being the decision-makers and often the main controllers of household resources. Thus they play a critical role in a woman's healthcare-seeking behaviour (Kaiser et al., 2019; Shaikh & Hatcher, 2004).

*Division of Labour.* Though the community shares in the object of the activity, the division of labour define their interaction and stipulate who can do what. This division is both horizontal, defining the responsibilities between members and vertical with regard to the division of power and status (Hsu et al., 2010; McMichael, 1999). In a healthcare-seeking context, various stakeholders have different responsibilities. However, in addition to these horizontal assignments, power dynamics may mediate how community members interact with the object of say a healthy pregnancy. For example, healthcare-seeking behaviour may be influenced by the health client's perception of the authority of health professionals. The health clients may therefore fear to engage the practitioners and feel compelled to conform to their 'directions' for fear of being labelled difficult (Entwistle et al., 2010;

Frosch et al., 2012). These patient-doctor relationships may influence how maternal health clients pursue their motive of a successful pregnancy.

Put together, Figure 1 illustrates the key concepts of AT as operationalised to define healthcare-seeking as an activity

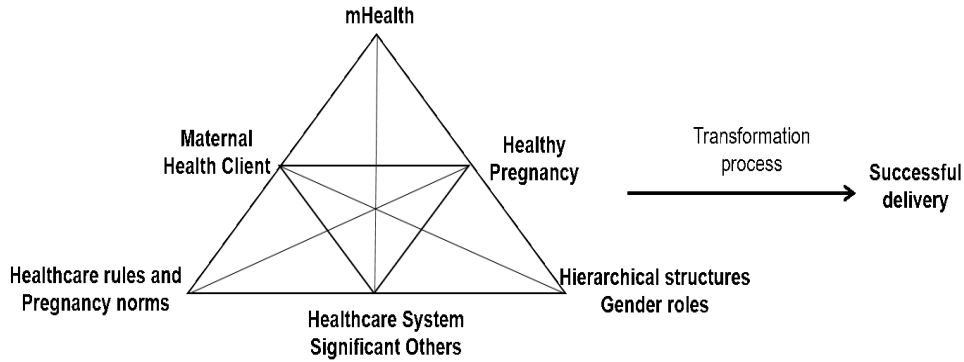

**Figure 1: A diagrammatic representation of maternal healthcare-seeking activity**

Other than the principles of unity of consciousness, object orientedness, and mediation that were mentioned earlier, AT provides three other powerful principles: the hierarchical structure of the activity, contradictions, and internalisation/externalisation; that may provide insight to the human-technology interaction within the activity context.

*The hierarchical structure of the activity* (Figure 2) posits that an activity which is the 'why' a subject engages in an endeavour like healthcare-seeking represents the top-most process layer. Next to this at a lower level are actions that encompass the 'what' is done to fulfil the motive of an activity, followed by operations that entail the 'how' people carry out these actions. While operations may be mental/cognitive processes that may not be observable depending on the study, it is reasonable to assume that the actions of the subject are manifestations of these subconscious processes.

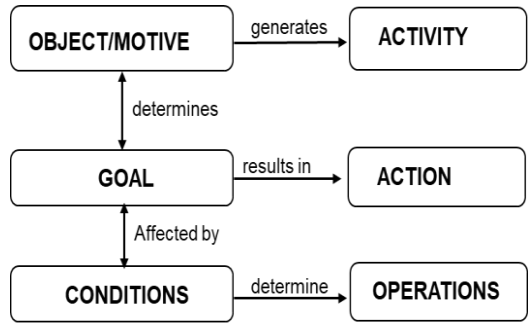

**Figure 2: Activity hierarchy showing what and how motives are pursued**

In a mHealth mediated healthcare-seeking context, therefore, the step to use mHealth may be considered an action within the activity hierarchy. This view is corroborated by Karanasios (2018). The author posits that in the IS context where an activity is mediated by an IT artifact, users' interaction with the technology can be viewed as an action. These levels are not fixed but can move up and down as conditions change (Kaptelinin & Nardi, 2006). Karanasios (2018) explains that when conditions (at the lowest level of the activity hierarchy) change, requiring actors to adopt new actions and goals, people may *adapt* provided the motive of the activity remains unchanged. However, people get unsatisfied when their motives are frustrated. Thus, the hierarchy of activity may prove useful if we are attempting to understand technology use in light of a subject's motives in a technology-mediated activity context. For example, how might users' interaction change in response to their use of technology enabling or constraining their desired motive? Do they adopt certain technology utilisation behaviours that may influence their long-term use? Such insights would prove useful in being able to explicate a 'continuum of use' from adoption to continuance which are important topics in IS acceptance.

*Contradictions* seen in structural tensions within the activity holds that the dialectical relationship between actors and their environments results in a type of development. In a healthcare-seeking context mediated by mHealth, we may analyse how these tensions develop within the activity, how the elements in the activity system (e.g., users/subjects) respond to these tensions and subsequently how such adaptations give further insights to technology use.

*Internalisation/Externalisation. Internalisation* is the internal reconstruction of an external operation (Allen et al., 2011). We posit that a researcher could further explicate technology use by analysing how the subject's internalisation of the IS artifact (mHealth) within the activity gives rise to varied utilisation behaviour. *Externalisation,* on the other hand, is the creation of either new artifacts, or ways of working, triggered by the awareness of tensions in the existing system (Frambach et al., 2014; Nowé, 2007). In light of the internal processes, *viz.* the subject's maternal socio-cultural context in maternal healthcare-seeking, and tensions raised by interacting with say mHealth, the maternal health clients may adopt new ways of using the mHealth intervention to solve existing tensions. The analysis of such externalisation allows the researcher to complete the dialectical study of the relationship between technology and the user as it allows one to explicate how the subjects' externalisation of their activity context shapes the use of technology. In other words, we argue that using AT, research questions that seek to explain the phenomenon of utilisation within a context of use may be phrased from the perspective of internalisation and externalisation to capture the complex relationship between users and technology.

To summarise this section, activity theory provides a set of concepts to unpack the relationship between technology and HIT utilization as elaborated in Table 1.

**Table 1: How conceptualization of healthcare-seeking as an activity may be applied to study HIT utilisation**

| Concept | Potential benefit | Sample operationalization |
|---|---|---|
| Activity | Provides the context for studying technology in use | The Healthcare-seeking activity |
| Subject | The individual (s) whose utilisation is being studied | The maternal health client |
| Object | Provides the means to understand not only 'what' activity (healthcare-seeking) that people are doing, but also 'why' they are doing it; thus, providing perspective to why health clients may use technology in the first place. | Pursuit of a successful pregnancy |
| Action | Actions can best gain meaning when examined under the context of the activity in which they are performed. As an action, the use of technology can best be explicated within the healthcare-seeking activity in which it is performed | Using mHealth technology during pregnancy with the goal to access pregnancy-related educational content |
| Activity Hierarchy | May be useful in understanding different 'subject' responses that may result from a change in goals and conditions within the activity. | Analysing how changes in conditions may subsequently affect the goals and motives of the healthcare-seeking activity and how such changes may affect actions such as the use of technology |
| Contradictions | Provides a means to analyse the healthcare-seeking activity for tensions that may influence/change the use | Explore how tensions impact the actions of the subject (technology user) |
| Internalisation / Externalisation | Provides an analytical means to study the dialectical relationship between technology and the subject more holistically. | Main study questions may be phrased as follows: How does the maternal health clients' internalisation of mHealth shape their utilisation? How does maternal health clients' externalisation of their healthcare-seeking socio-cultural context shape their mHealth use? |

### 2.3 Opportunities for AT in Studying Technology Use

Though use has traditionally been defined as involving a user, employing an object (system) to accomplish a task (Burton-Jones & Straub Jr, 2006), individuals' usage decisions happen within rich social contexts (Lewis et al., 2003) that would provide far richer understanding if included in the study of the phenomenon. Newer definitions (e.g. Barki et al., 2007) encompass a person's interaction with technology where a user employs IT to accomplish a certain task, as well as the person's activities to adapt, change or modify the human-technology interaction context. Two benefits of AT to study use stand out; i) the relationship between the subject, tools and objects as purposeful activities in AT is in harmony to the traditional construction of 'use'; ii) owing to its ability to incorporate an structure/activity and context in the same environment, AT is able provide a more nuanced understanding of usage decisions influenced by context.

We will analyse one empirical case using the operationalisation presented in Table 1. The perspective of healthcare-seeking as an activity will be used to illustrate the technology use phenomena and how this perspective may provide a richer means to reflect on technology design and implementation implications.

## 3. METHODOLOGY

The subset of research presented here constitutes data gained from a research project whose data was collected between January and May 2019. The project adopted a case study design and was qualitative in nature, using interviews, focus group discussions and participant observations to collect data. A total of 38 interviews that were between 45 – 60 minutes were conducted, including maternal health clients, their partners and key informants involved in the implementation of the mHealth project. The women were largely of low socio-economic status and at least 75% being between the ages of 20 – 29 years. Almost all the women reported being married and less than one third had attained a tertiary level of education. Three focus group meetings, lasting approximately 90 minutes each were organised: two with 14 maternal health clients and one with the participating partners. All primary data from the interviews and FGDs were translated and transcribed verbatim into English. The transcripts were uploaded to NVivo v12 for further analysis. The analysis followed a hybrid approach entailing both deductive and inductive analyses. The use of Braun and Clarke's (2006) thematic analysis allowed themes to emerge from data and the theoretical framework (Frambach, Driessen, & van der Vleuten, 2014). As such, using the principles of AT as a guide, the initial codes to represent the data categories were developed based on the concepts of the theory. Thereafter, these categories of data were inductively analysed to develop further themes. Thus, while the lower-level themes were represented by AT concepts such as object, mediation, rules etc., higher-level themes were inductively developed from the prominent themes that emerged from these lower-level categories of the data. The insights from the data analysis are used to both enrich the description of the context; since in AT the context is in fact the activity system itself (Nardi, 1996b), as well as offer theoretical insights.

## 4. THE MATERNAL HEALTHCARE SEEKING CONTEXT

This section describes the PROMPTS intervention, a mobile phone service in Kenya that was implemented as a way to send stage-appropriate pregnancy-related messages to maternal health clients from low socio-economic groups living in peri-urban areas of the capital city Nairobi. The mHealth intervention was implemented as a toll-free service. Automated messaging was combined with a clinician-supported helpdesk to answer maternal health clients' questions. Upon interacting with the mHealth poster at the ANC clinic, interested mothers could self-register by sending a short message to a short message code. The mother was then required to respond to a short SMS survey where they provided details of their gestation age and their preferred language for receiving the health messages. The SMS service did not require the women to provide any personal information like name or residential address.

Such mHealth services are prevalent in Kenya as in most developing countries where mHealth holds the potential to help developing countries to achieve their health-related developmental goals. The health sector conditions in such countries that make mHealth a promising technology for transforming the health sector include distance of health facilities, high cost of care and a resource-strained healthcare system to mention a few (Gabrysch & Campbell, 2009; Mannava et al., 2015; Simkhada et al., 2008).

However, while healthcare demand and supply factors influence when, why and how health clients use health interventions, other reasons like socio-cultural factors shape healthcare-seeking behaviour. This is particularly evident in maternal healthcare-seeking where numerous pregnancy beliefs, norms and values may inform perceptions about healthcare-seeking. In this empirical context, like in other developing country contexts (see Mumtaz & Salway, 2007; Zamawe, 2013), pregnancy was seen as a normal condition, with some of the potential risks being considered a natural part of the journey. Consequently, the women sought care only when they experienced something that was deemed to be unusual or when a symptom persisted. The women believed that the appropriate time to start ANC was when the pregnancy begins to 'show', and this was mainly attributed to the need to protect the pregnancy from any undesired eventualities like witchcraft. *"Some people are not good, and they might do some things that might cause you to lose the pregnancy through miscarriage"* [Mother 4].

To be admitted in a healthcare facility for delivery, mothers in the empirical context needed to have a clinic card, a failure to which they may have risked being turned back by the healthcare workers when they were due for delivery. This is a common predicament that maternal health clients in other developing country contexts face (Finlayson & Downe, 2013; Mrisho et al., 2007).

## 5. ANALYSIS

In this section, we demonstrate the application of the theoretical framework that was discussed in the previous section to analyse the PROMPTS empirical context. The analysis reveals that the desire for a healthy pregnancy, both directed and motivated the maternal health clients' healthcare-seeking activity, while mHealth as a technology artifact mediated the healthcare-seeking activity as shown in Table 2. Consequently, the action of using the mHealth intervention was to pursue goals like accessing pregnancy-related care and support, and pregnancy information, within the realities and conditions of the existing healthcare-seeking context such as a resource-constrained healthcare system.

**Table 2: How the use of AT facilitated an understanding of technology mediation in its context of use**

| Concept | Empirical scenario | Analysis enabled by AT | |
|---|---|---|---|
| Community (of Purpose) and Division of Labour | Courtesy of their position as care providers, the HCPs were expected to provide information, care, and support. <br><br> Older female kin courtesy of their experience were expected to support the maternal health client. | Analysis of tensions between elements of the activity, in this case, the CoP and Division of labour | mHealth provided an avenue to resolve this contradiction by providing alternative access to pregnancy-related information and care resulting in partial substitution of the CoP. |
| Rules/Values | Cultural norms restricting maternal health clients to seek maternal healthcare services only when the pregnancy was 'showing' | An analysis of how IT can support rules in an activity *viz*. imposing new rules, making them visible, or negotiating rules (Nardi, 1996a) | The mHealth intervention allowed women to interact with maternal services while maintaining their value system. |

### 5.1 Conceptualising use as internalisation

*Adoption of technology.* While some uncertainties facilitated the adoption of mHealth, some resulted in hesitation to adopt the technology. Using the intervention proved useful in mitigating the pregnancy-related uncertainties associated with openly sharing one's pregnancy for the fear of an 'evil eye'/witchcraft. However, while it presented the opportunity to access useful information to mitigate uncertainties related to endangering the pregnancy from lack of/following wrong information, the use of the intervention also presented concerns on the trustworthiness of the information with regards to harming the pregnancy.

Other uncertainties related to the material context influenced the maternal health clients' response to the intervention. The popularly used money transfer service 'M-Pesa', also gave opportunities for fraudulent activities leading unsuspecting subscribers to loose money in social engineering scams (Buku & Meredith, 2012). Thus, mobile phone users were often suspicious of unsolicited messages. The study participants were suspicious of the intervention for fear that it was a scam that would lead to their loss of airtime. *"They might say it's free and maybe you have only twenty Shillings' credit[1], and then you send the message, and they [the mHealth intervention] consume your credit."* [FGD Participant]. This uncertainty led to hesitation to adopt the intervention.

Learning about the intervention from the health facility, therefore seemed to engender a sense of trust that the intervention was legitimate while the features like the toll-free design enabled the maternal health clients to overcome initial perceptions of risk and uncertainty about the intervention being a fraud. Being a toll-free service may have also increased the trialability of the intervention and afforded the women the agency to make mHealth adoption-related decisions. The trialability of an innovation is positively correlated to the likelihood of adoption (Rogers, 2010). These insights in the analysis were useful in seeing how technology characteristics and implementation circumstances may have engendered initial trust. The initial trust subsequently contributed to helping technology users overcome their perceptions of risk and uncertainty associated with the use of new technology. This made the following proposition possible:

> *Proposition 1. Where context and technology characteristics provide sufficient resources for the artifact to be sufficiently legitimised for initial trust, users are likely to adopt new technology.*

*Dissonance from using technology as a substitute.* The action of using mHealth to pursue a healthy pregnancy was made possible by the existing conditions presented in the healthcare-seeking context *viz.* a resource-constrained healthcare system, negative attitudes and behaviour of clinicians, and socioeconomic barriers which affected when and how the maternal clients could use traditional healthcare services. The complexity and tensions between the community and division of labour (Tensions 3, 4 in Figure 3) resulted in further challenges in the healthcare-seeking activity. While it was the 'responsibility' of clinicians and next of kin to provide pregnancy-related care and education, this division of labour failed because of:

1. The spatial separation between the women and their experienced, older relatives whose responsibility it was to offer information and support.
2. The low doctor-patient ratio which resulted in hurried consultations with limited time to provide sufficient information. *"It is often difficult to give individualised care because of the workload"* [Healthcare Providers].

All these factors led to a perceived usefulness of mHealth for the maternal health clients to overcome these barriers, thus leading to the use of mHealth as a substitute. Using the intervention for a purpose other than that which it was designed for (undesired technology framing) led to incongruence and further tensions (depicted as 1 in Figure 3). The foregoing analysis enabled us to suggest the following proposition:

> *Proposition 2: When the perceived usefulness of mHealth presents the opportunity to use it as a substitute, mHealth users may develop undesired technology framing and experience dissonance.*

### 5.2 Conceptualising use as externalisation

As maternal health clients experienced tensions/contradictions/challenges within the activity system (Figure 3), further use behaviour was elicited in the attempt to resolve the tensions. We discuss a few of the possibilities for understanding use that was gained from the analysis of some of these contradictions.

---

[1] Prepaid airtime or talktime that allows the mobile phone user to make calls or send text messages.

*Legitimisation.* This understanding is obtained by analysing contradiction 6 in Figure 3. At the core of legitimacy is reducing uncertainty by developing trust (Deephouse & Suchman, 2008). Our analysis showed that maternal health clients legitimised the mHealth intervention to build further (post-adoptive) trust. This was achieved by using the characteristics of the intervention and messages being shared via the platform, as well as comparisons with other sources of information to draw 'judgements' on the trustworthiness of both the information and the intervention itself (Sowon & Chigona, 2020). The characteristics of the intervention such as its responsiveness, anonymity, facilitated the legitimisation of the intervention beyond initial trust. Post adoptive trust is necessary for use continuance (Akter et al., 2013; Kaium et al., 2020). The analysis made the following proposition possible:

> **Proposition 3:** *mHealth users employ technology characteristics and the community of purpose to legitimise the intervention beyond initial trust*

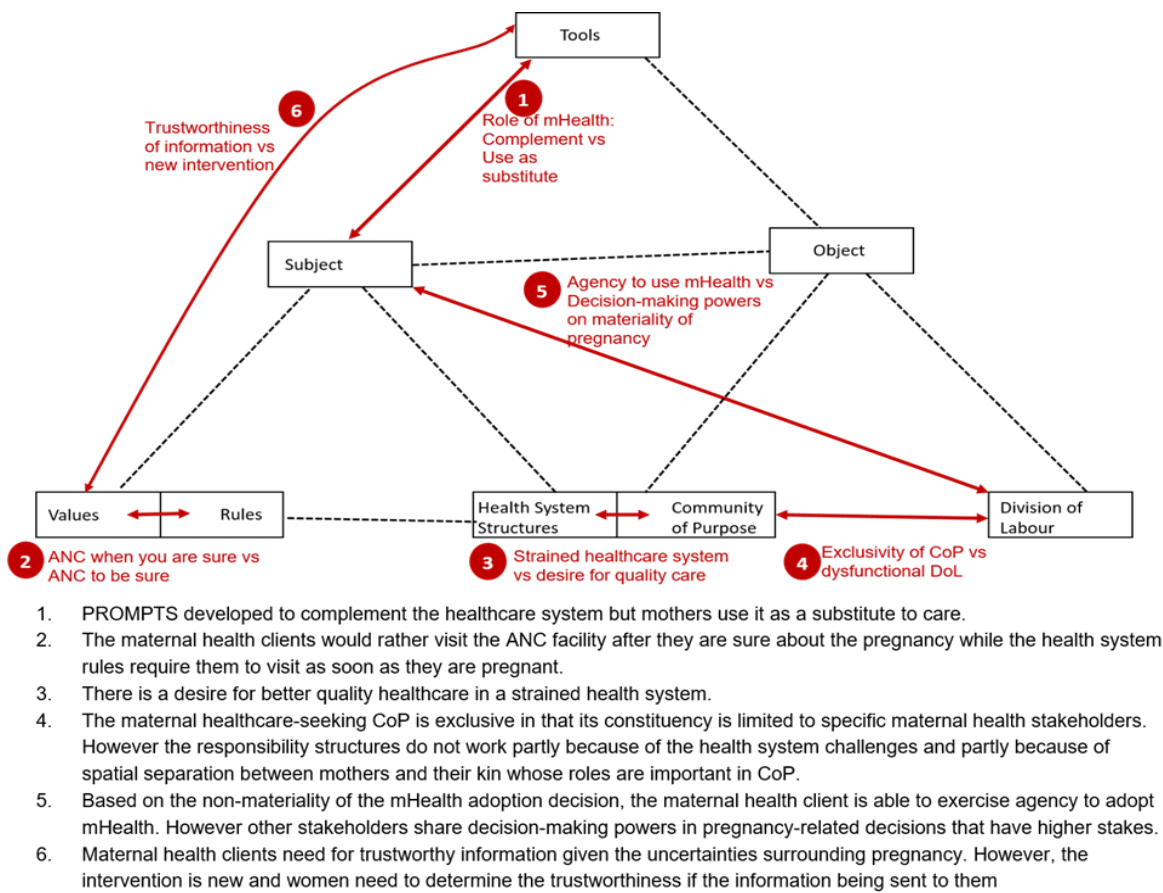

**Figure 3: Contradictions in the maternal healthcare-seeking activity system**

*Negotiated use.* While the design features of the intervention, as well as the components of the material context, may have facilitated agency to adopt, further analysis of contradiction 5 (Figure 3) revealed a change in dynamics when it came to appropriating the pregnancy-related information that was received through the intervention. So long as the intervention was offered free of charge and was associated with a legitimate entity, the decision to adopt mHealth may have been immaterial in terms of consequences. However, since the appropriation of information may have reflected a more material decision in the CoP reflecting a higher risk to the shared motive of a successful pregnancy, maternal health clients engaged in a process of negotiating their mHealth use. Their aim here was to make their use of mHealth a culturally appropriate behaviour given the interdependent nature of the maternal healthcare-seeking context in the empirical setting. The idea of seeking appropriate use has been alluded to before by other theorists like Rogers (1995) in concepts like *compatibility* in the diffusion of innovations theory. Following this analysis, the following proposition was possible:

> **Proposition 4:** *When stakeholders play a vital role in the CoP, mHealth users employ legitimated mHealth interventions to negotiate use.*

*Technology coping strategies.* As noted earlier, the varied use of technology in a manner different from what it was intended for resulted in dissonance for the maternal health clients. This mainly occurred in situations when the intervention 'failed' in engaging them by providing quick feedback, whenever the mothers were using it to access urgent care. "*I thought [the delay] was not good because it could be an emergency and sometimes when asking for information you require the response immediately...*". To resolve this, further analysis of the contradiction indicates that the mothers adopted varied technology coping strategies. Some mHealth users accommodated the technology, whereas some completely abandoned it.

AT's principle of the hierarchy of activity was useful to further elucidating the difference in behaviour. Some analysis using this perspective indicated that where the mHealth intervention was sufficiently legitimised for the maternal health client, the need to wait for a response reflected a change in conditions rather than the frustration of the maternal healthcare-seeking motive. This led to adopting technology accommodation coping strategies. On the other hand, for the maternal health clients for whom the mHealth technology had not been sufficiently legitimised, the delay in receiving a response threatened their very motive of healthcare-seeking. For this reason, such users abandoned the use of the intervention. The analysis, therefore, allowed us to make the proposition that:

> **Proposition 5:** *Health clients may abandon technology (e.g., mHealth) when it is deemed to frustrate their healthcare-seeking motive.*

## 6. DISCUSSION AND CONCLUSION

This study was founded on the argument that studying technology in situ is a much better perspective to studying concepts of human technology interaction like use. The study thus proceeded to demonstrate how conceptualising healthcare-seeking as an activity may prove useful to this end. Traditionally, technology use has been explicated using theories like TAM and UTAUT. Unlike these theories, we have demonstrated that AT allows for a broader understanding of use while not losing the original definition that entails a user, system and task. (Burton-Jones & Straub Jr, 2006). However, since individuals' usage decisions happen within rich social contexts (Lewis et al., 2003), we explained that broader perspectives would provide a far richer understanding. Hence, rather than just conceptualising use as a user employing IT to accomplish a certain task, we approached it as including a person's activities to adapt, change or modify the human-technology interaction context. In this study, we have demonstrated that AT has all the necessary vocabulary and conceptual underpinning to extensively explicate an individual's activities to adapt, change or modify the technology, thus offering richer insights on use. Such perspectives may be what the IS field needs to provide a 'next-level' understanding of the interaction between technology and humans within their social contexts. It is evident that such an expanded view may better explain salient technology acceptance outcomes without precluding IS use behaviours included in the traditional view of the construct. It also helps us to move beyond adoption to understanding the complete continuum of user-technology interaction in a way that allows researchers to reflect on the design and implementation implications.

Adopting an activity perspective to understanding user-technology interaction especially in a context such as mHealth where the artifact is a combination of the technology itself and the information may complicate the idea of use and the understanding of the user technology interaction at the two levels. However, analytically separating the IT construct as we have done here allows for richer insights to emerge. For example, while use of mHealth as a technology was an autonomous decision, the appropriation of information depended on the material context which necessitated the maternal health clients to negotiate use.

We conclude by suggesting that AT offers a powerful tool to investigate the situated entanglement without embracing technological determinism, taking the technology for granted, or allowing the technology to vanish from view (Orlikowski, 2005). The scarcity of AT application may be associated with its robust, complex, and abstract nature (Mwanza, 2001; Wiser et al., 2019). The application of AT requires the researcher to select a practical level of abstraction (Allen et al., 2011; Engeström, 2001). We believe that our application is an appropriate attempt to operationalise AT in Information Systems especially as a tool that can be used to understand use. The key concepts can be applied as sensitising devices in analysing empirical cases. The analysis of mHealth within the context of the healthcare-seeking activity not only explicated the phenomenon of use but was also useful in revealing the different design and implementation characteristics of the context and the technology such as were useful in shaping the maternal health clients' utilisation behaviour. Such insights could provide a patient-centred perspective to the design and implementation of health information systems. Thus, the benefits of conceptualising healthcare-seeking as an activity serve an important role in bringing together the HCI and IS fields. These have historically been known to adopt varied theoretical perspectives according to their focal interest: the former on design of technology artefacts, and the latter on IS-related development and change (Avgerou, 2010).

This study uses one human engagement (healthcare-seeking) to operationalise AT and to show how the phenomenon of use may be understood. Insights on use may vary in other human activities where technology is used. We, therefore, encourage other IS researchers to test the application of AT presented here, in other technology contexts. This will serve to validate the findings that have been presented here and to build further evidence that would be useful in the further theorization of technology utilisation phenomena while building the evidence on how AT may be operationalised in IS research.

On a practical and theoretical level, the complexity of certain AT concepts may be more complicated than could be captured in this study. For example, the dichotomy between internalisation and externalisation, when one ends and where the other begins may limit the extent to which they can be separated for analytical accessibility. The conceptualisation we presented here may also be limited to the extent that we adopt an instantiation of the activity at a particular point in time. We encourage other researchers to explore the value of the activity system and the aspect of a changing activity based on tensions, contradictions and changing needs and motives of the subject on explicating utilisation.

## 7. ACKNOWLEDGEMENT

**REFERENCES**

Aikins, A. de-Graft. (2014). Food Beliefs and Practices During Pregnancy in Ghana: Implications for Maternal Health Interventions. *Health Care for Women International*. https://www.tandfonline.com/doi/abs/10.1080/07399332.2014.926902

Akter, S., D'Ambra, J., Ray, P., & Hani, U. (2013). Modelling the impact of mHealth service quality on satisfaction, continuance and quality of life. *Behaviour & Information Technology*, *32*(12), 1225–1241.

Alexander, J. A., & Hearld, L. R. (2012). Methods and metrics challenges of delivery-system research. *Implementation Science*, *7*(1), 15.

Allen, D., Brown, A., Karanasios, S., & Norman, A. (2013). How Should Technology-Mediated Organizational Change Be Explained? A Comparison of the Contributions of Critical Realism and Activity Theory. *Mis Quarterly*, *37*(3), 835–854.

Allen, D., Karanasios, S., & Slavova, M. (2011). Working with activity theory: Context, technology, and information behavior. *Journal of the Association for Information Science and Technology*, *62*(4), 776–788.

Avgerou, C. (2010). Discourses on ICT and development. *Information Technologies & International Development*, *6*(3), 1–18.

Awa, H. O., Ukoha, O., & Emecheta, B. C. (2016). Using TOE theoretical framework to study the adoption of ERP solution. *Cogent Business & Management*, *3*(1), 1196571.

Bannon, L. J., & Bødker, S. (1989). Beyond the interface: Encountering artifacts in use. *DAIMI Report Series*, *288*.

Barki, H., Titah, R., & Boffo, C. (2007). Information system use–related activity: an expanded behavioral conceptualization of individual-level information system use. *Information Systems Research*, *18*(2), 173–192.

Benson, A., Lawler, C., & Whitworth, A. (2008). Rules, roles and tools: Activity theory and the comparative study of e-learning. *British Journal of Educational Technology*, *39*(3), 456–467.

Blayone, T. J. (2019). Theorising effective uses of digital technology with activity theory. *Technology, Pedagogy and Education*, 1–16.

Bødker, S. (1996). Applying activity theory to video analysis: how to make sense of video data in HCI. *Context and Consciousness: Activity Theory and Human Computer Interaction*, 147–174.

Braun, V., & Clarke, V. (2006). Using thematic analysis in psychology. *Qualitative Research in Psychology*, *3*(2), 77–101. https://doi.org/10.1191/1478088706qp063oa

Buku, M. W., & Meredith, M. W. (2012). Safaricom and M-Pesa in Kenya: financial inclusion and financial integrity. *Wash. Jl Tech. & Arts*, *8*, 375.

Burton-Jones, A., & Straub Jr, D. W. (2006). Reconceptualizing system usage: An approach and empirical test. *Information Systems Research*, *17*(3), 228–246.

Cecez-Kecmanovic, D., Galliers, R. D., Henfridsson, O., Newell, S., & Vidgen, R. (2014). The Sociomateriality of Information Systems. *Mis Quarterly*, *38*(3), 809–830.

Engeström, Y. (2001). Expansive Learning at Work: Toward an activity theoretical reconceptualization. *Journal of Education and Work*, *14*(1), 133–156. https://doi.org/10.1080/13639080020028747

Ensor, T., & Cooper, S. (2004). Overcoming barriers to health service access: influencing the demand side. *Health Policy and Planning*, *19*(2), 69–79.

Entwistle, V. A., McCaughan, D., Watt, I. S., Birks, Y., Hall, J., Peat, M., Williams, B., & Wright, J. (2010). Speaking up about safety concerns: multi-setting qualitative study of patients' views and experiences. *Qual Saf Health Care*, *19*(6), e33–e33.

Finlayson, K., & Downe, S. (2013). Why do women not use antenatal services in low-and middle-income countries? A meta-synthesis of qualitative studies. *PLoS Medicine*, *10*(1).

Frambach, J. M., Driessen, E. W., & van der Vleuten, C. P. M. (2014). Using activity theory to study cultural complexity in medical education. *Perspectives on Medical Education*, *3*(3), 190–203. https://doi.org/10.1007/s40037-014-0114-3

Frosch, D. L., May, S. G., Rendle, K. A., Tietbohl, C., & Elwyn, G. (2012). Authoritarian physicians and patients' fear of being labeled 'difficult' among key obstacles to shared decision making. *Health Affairs*, *31*(5), 1030–1038.

Gabrysch, S., & Campbell, O. M. (2009). Still too far to walk: literature review of the determinants of delivery service use. *BMC Pregnancy and Childbirth*, *9*(1), 34.

Hautasaari, A. (2013). 'Could someone please translate this?' activity analysis of wikipedia article translation by non-experts. *Proceedings of the 2013 Conference on Computer Supported Cooperative Work*, 945–954.

Hsu, P.-L., van Eijck, M., & Roth, W.-M. (2010). Students' representations of scientific practice during a science internship: Reflections from an activity-theoretic perspective. *International Journal of Science Education*, *32*(9), 1243–1266.

Kaiser, J. L., Fong, R. M., Hamer, D. H., Biemba, G., Ngoma, T., Tusing, B., & Scott, N. A. (2019). How a woman's interpersonal relationships can delay care-seeking and access during the maternity period in rural Zambia: An intersection of the Social Ecological Model with the Three Delays Framework. *Social Science & Medicine*, *220*, 312–321. https://doi.org/10.1016/j.socscimed.2018.11.011

Kaium, M. A., Bao, Y., Alam, M. Z., & Hoque, M. R. (2020). Understanding continuance usage intention of mHealth in a developing country. *International Journal of Pharmaceutical and Healthcare Marketing*.

Kaptelinin, V., & Nardi, B. A. (2006). *Acting with technology: Activity theory and interaction design*. MIT press.

Karakus, T. (2014). Practices and potential of activity theory for educational technology research. In *Handbook of research on educational communications and technology* (pp. 151–160). Springer.

Karanasios, S. (2018). Toward a unified view of technology and activity: The contribution of activity theory to information systems research. *Information Technology & People*, *31*(1), 134–155. https://doi.org/10.1108/ITP-04-2016-0074

Karanasios, S., & Allen, D. (2014). Mobile technology in mobile work: contradictions and congruencies in activity systems. *European Journal of Information Systems*, *23*(5), 529–542.

Klein, H. K., & Myers, M. D. (2001). A classification scheme for interpretive research in information systems. In *Qualitative research in IS: issues and trends* (pp. 218–239). IGI Global.

Kuutti, K. (1996). Activity theory as a potential framework for human-computer interaction research. *Context and Consciousness: Activity Theory and Human-Computer Interaction*, *1744*(Journal Article).

Lee, S. H., Nurmatov, U. B., Nwaru, B. I., Mukherjee, M., Grant, L., & Pagliari, C. (2016). Effectiveness of mHealth interventions for maternal, newborn and child health in low–and middle–income countries: Systematic review and meta–analysis. *Journal of Global Health*, *6*(1).

Leonardi, P. M. (2011). When flexible routines meet flexible technologies: Affordance, constraint, and the imbrication of human and material agencies. *MIS Quarterly*, 147–167.

Leontiev, A. N. (1978). *Activity, Consciousness, and Personality*. Prentice-Hall Englewood Cliffs, Nj.

Lewis, W., Agarwal, R., & Sambamurthy, V. (2003). Sources of influence on beliefs about information technology use: An empirical study of knowledge workers. *MIS Quarterly*, 657–678.

Mannava, P., Durrant, K., Fisher, J., Chersich, M., & Luchters, S. (2015). Attitudes and behaviours of maternal health care providers in interactions with clients: a systematic review. *Globalization and Health*, *11*(1), 36.

McMichael, H. (1999). *An activity-based perspective for information systems research*. 10th Amsterdam Conference on Information Systems. In proceedings.

Mrisho, M., Schellenberg, J. A., Mushi, A. K., Obrist, B., Mshinda, H., Tanner, M., & Schellenberg, D. (2007). Factors affecting home delivery in rural Tanzania. *Tropical Medicine & International Health*, *12*(7), 862–872.

Mumtaz, Z., & Salway, S. M. (2007). Gender, pregnancy and the uptake of antenatal care services in Pakistan. *Sociology of Health & Illness*, *29*(1), 1–26.

Mursu, A. S., Luukkonen, I., Toivanen, M., & Korpela, M. J. (2006). Activity Theory in information systems research and practice: theoretical underpinnings for an information systems development model. *Information Research*, *12*(3), 3.

Mwanza, D. (2001). *Where theory meets practice: A case for an activity theory based methodology to guide computer system design*.

Nardi, B. A. (1996a). Activity theory and human-computer interaction. *Context and Consciousness: Activity Theory and Human-Computer Interaction*, *436*, 7–16.

Nardi, B. A. (1996b). Studying context: A comparison of activity theory, situated action models, and distributed cognition. *Context and Consciousness: Activity Theory and Human-Computer Interaction*, Journal Article, 69–102.

Nowé, K. (2007). *Tensions and Contradictions in Information Management*. Department of Library and Information Science/Swedish School of Library and ….

Nyemba-Mudenda, M., & Chigona, W. (2018). mHealth outcomes for pregnant mothers in Malawi: a capability perspective. *Information Technology for Development*, *24*(2), 245–278. https://doi.org/10.1080/02681102.2017.1397594

Olenja, J. (2003). Editorial Health seeking behaviour in context. *East African Medical Journal*, *80*(2), 61–62.

Orlikowski, W. J. (2005). Material works: Exploring the situated entanglement of technological performativity and human agency. *Scandinavian Journal of Information Systems*, *17*(1), 6.

Pell, C., Meñaca, A., Were, F., Afrah, N. A., Chatio, S., Manda-Taylor, L., Hamel, M. J., Hodgson, A., Tagbor, H., & Kalilani, L. (2013). Factors affecting antenatal care attendance: results from qualitative studies in Ghana, Kenya and Malawi. *PloS One*, *8*(1).

Rogers, E. M. (1995). *Diffusion of Innovations, by Everett Rogers (1995)*. http://www.stanford.edu/class/symbsys205/Diffusion%20of%20Innovations.htm

Shaikh, B. T., & Hatcher, J. (2004). Health seeking behaviour and health service utilization in Pakistan: challenging the policy makers. *Journal of Public Health*, *27*(1), 49–54.

Simkhada, B., Teijlingen, E. R. van, Porter, M., & Simkhada, P. (2008). Factors affecting the utilization of antenatal care in developing countries: systematic review of the literature. *Journal of Advanced Nursing*, *61*(3), 244–260.

Sowon, K., & Chigona, W. (2020). Trust in mHealth: How do Maternal Health Clients Accept and Use mHealth Interventions? *Conference of the South African Institute of Computer Scientists and Information Technologists 2020*, 189–197.

Uldbjerg, C. S., Schramm, S., Kaducu, F. O., Ovuga, E., & Sodemann, M. (2020). Perceived barriers to utilization of antenatal care services in northern Uganda: A qualitative study. *Sexual & Reproductive Healthcare*, *23*, 100464. https://doi.org/10.1016/j.srhc.2019.100464

Venkatesh, V., Davis, F., & Morris, M. G. (2007). Dead or alive? The development, trajectory and future of technology adoption research. *Journal of the Association for Information Systems*, *8*(4), 1.

Wang, W., Alva, S., Wang, S., & Fort, A. (2011). *Levels and trends in the use of maternal health services in developing countries*. Calverton Maryland ICF Macro MEASURE DHS 2011 Jun.

Wiser, F., Durst, C., & Wickramasinghe, N. (2019). Using activity theory successfully in healthcare: A systematic review of the theory's key challenges to date. *Proceedings of the 52nd Hawaii International Conference on System Sciences*.

Yamagata-Lynch, L. C. (2010). *Activity Systems Analysis Methods: Understanding Complex Learning Environments*. Springer Science & Business Media.

Zamawe, C. O. (2013). Factors that affect maternal care seeking behaviour and the choice of practitioner (s) during complications: The case of Mang'anja tribe in Malawi. *Research on Humanities and Social Sciences*, *3*(18), 18–26.

Zurita, G., & Nussbaum, M. (2007). A conceptual framework based on activity theory for mobile CSCL. *British Journal of Educational Technology*, *38*(2), 211–235.